\documentstyle[amssymb,aps,prc,eqsecnum,12pt]{revtex}
\tightenlines

\begin{document}
\draft
\title{Radiative Corrections to Elastic Electron-Proton Scattering for Polarized
Electrons}
\author{L.C. Maximon and W.C. Parke}
\address{Physics Department\\
The George Washington University\\
Washington, DC 20052}
\date{22 November 1999}
\maketitle

\begin{abstract}
We analyze the radiative correction to high energy elastic electron-proton
scattering of polarized electrons. We show that if the approximations
inherent in the calculations developed by Tsai and given in the work of Mo
and Tsai, which have been used in the analysis of almost all experimental
data pertaining to medium and high energy elastic electron scattering for
the past three decades, are maintained, then the same radiative correction
applies both in the case of initially polarized and unpolarized electrons.
\end{abstract}

\pacs{13.40.K, 25.30.Bf, 13.60.Fz}

\section{Introduction}

The radiative correction to elastic electron-proton scattering is well-known
from the work of Tsai \cite{tsai} and Mo and Tsai \cite{mo}, and the
expressions given in \cite{mo} have been used in the analysis of almost all
experimental data pertaining to medium and high energy elastic electron
scattering for the past three decades. Very recently, experiments using
polarized electron beams have been carried out at Jefferson Lab \cite{HallA}%
; specifically, longitudinally polarized electrons were scattered from
unpolarized protons ($\overrightarrow{e}p\rightarrow e\overrightarrow{p}$)
and the transverse and longitudinal polarizations of the recoil protons were
measured in order to obtain the ratio of the proton's elastic
electromagnetic form factors, $G_{E_{p}}/G_{M_{p}}$. Given that radiative
corrections to elastic electron-proton scattering are generally of the order
of 20\% - 30\% for four-momentum transfer squared in the range considered in
these experiments (0.5 to 3.5 (GeV/c)$^{2}$), the question arises as to
whether the same radiative correction used in the case of unpolarized beams
and targets can be applied in the case of polarized electron beams when the
polarization of the recoil proton is measured. We show here that if the
approximations inherent in the calculations developed in \cite{tsai} and
given in \cite{mo} are maintained, then the same radiative correction
applies both in the case of initially polarized and unpolarized electrons.
In Sec. II we present the cross section for the scattering of polarized
electrons from unpolarized protons in the absence of radiative corrections.
In Sec. III we give each of the matrix elements associated with the
radiative correction and discuss the significant approximations that are
made in \cite{tsai} to evaluate their contribution to the cross section. We
then show that with these approximations the radiative corrections do not
depend on the polarization of either the electron or the proton in the
initial or final state.

\section{Differential cross section for scattering of polarized electrons}

\smallskip

The differential cross section for the scattering of polarized electrons
from unpolarized protons can be derived using standard techniques of quantum
electrodynamics. We follow the conventions of Bjorken and Drell \cite{bj};
the metric is defined by $p_{i}\cdot p_{j}=\epsilon _{i}\epsilon _{j}-{\bf p}%
_{i}\cdot {\bf p}_{j}$. Further, $\alpha =e^{2}/4\pi =1/137.036$; $m$ is the
electron rest mass; $M$ is the target nucleus rest mass; $\kappa $ the
anomalous magnetic moment of the proton; $p_{1}$ and $p_{3}$ the initial and
final electron four-momenta, respectively; $p_{2}$ and $p_{4}$ the initial
and final target nucleus four-momenta, respectively; $%
q=p_{1}-p_{3}=p_{4}-p_{2}$ is the four-momentum transfer to the target
nucleus for elastic scattering.

For one-photon exchange, the matrix element is 
\begin{equation}
M_{0}=e^{2}\overline{u}(p_{3})\gamma ^{\mu }u(p_{1})\frac{(-i)}{%
q^{2}+i\epsilon }\overline{u}(p_{4})\Gamma _{\mu }(q^{2})u(p_{2}),
\label{M0}
\end{equation}
whose magnitude squared, summed over final electron spin and averaged over
initial proton spin, is 
\begin{equation}
\left| \overline{M}_{0}\right| ^{2}=\frac{1}{2}\text{Tr}\left\{ \gamma ^{\nu
}\Lambda _{3}\gamma ^{\mu }\Lambda _{1}\Sigma _{1}\right\} \text{Tr}\left\{
\Sigma _{4}\Lambda _{4}\Gamma _{\mu }\Lambda _{2}\widetilde{\Gamma }_{\nu
}\right\} ,  \label{matrix}
\end{equation}
where $\Lambda _{i}=\left( p\!\!\!/_{i}+m_{i}\right) /\left( 2m_{i}\right) $
and $\Sigma _{i}=\left( 1+\gamma _{5}s\!\!\!/_{i}\right) /2$ are energy and
spin projection operators and 
\begin{equation}
\Gamma _{\mu }=F_{1}(q^{2})\gamma _{\mu }+\kappa F_{2}(q^{2})\frac{i\sigma
_{\mu \alpha }q^{\alpha }}{2M},\text{ }(\widetilde{\Gamma }_{\nu }\equiv
\gamma ^{0}\Gamma _{\nu }^{\dagger }\gamma ^{0})  \label{currop}
\end{equation}
is the proton-current operator. We assume high energies for the initial and
final electrons ($\epsilon _{1},\epsilon _{3}>>m$) and large momentum
transfers ($-q^{2}>>m^{2}$). Further, we express the cross section in terms
of the Sachs form factors, $G_{E}(q^{2})$ and $G_{M}(q^{2})$, which are
defined in terms of $F_{1}$ and $F_{2}$ by 
\begin{equation}
G_{E}=F_{1}-\tau \kappa F_{2},\,\,\,\,\,G_{M}=F_{1}+\kappa F_{2}  \label{sff}
\end{equation}
where $\tau =-q^{2}/4M^{2}$. Finally, we express the spin polarization
four-vectors of the initial electron and final proton, $s_{1}$ and $s_{4}$
respectively, in terms of the three-dimensional unit vectors specifying the
spin direction of the particles in their respective rest frames, ${%
\bbox{\zeta }}_{1}$ and ${\bbox{\zeta }}_{4}$. In general, for a particle of
mass $m$, and four-momentum $p=(\epsilon ,{\bf p})$, the four-vector $s$ is
given in terms of ${\bbox{\zeta }}${\bf \ }by \cite{olsen}, \cite{berest}
\begin{eqnarray}
s_{0} &=&\frac{{\bbox{\zeta }}\cdot {\bf p}}{m}  \label{spin} \\
{\bf s} &=&{\bbox{\zeta }}+{\bf p}\left[ \frac{{\bbox{\zeta }}\cdot {\bf p}}{%
m(m+\epsilon )}\right] .  \nonumber
\end{eqnarray}
For the initial electron we have, neglecting terms of relative order $%
m/\epsilon _{1}$, 
\begin{equation}
s_{1}\doteq hp_{1}/m  \label{espin}
\end{equation}
where $h\equiv {\bbox{\zeta }}_{1}\cdot \widehat{{\bf p}}_{1}$. The cross
section for the scattering of high energy polarized electrons into the
direction $\theta $ by unpolarized protons initially at rest is then 
\begin{eqnarray}
\frac{d\sigma }{d\Omega } &=&\frac{\alpha ^{2}\epsilon _{3}\cos ^{2}\frac{%
\theta }{2}}{4\epsilon _{1}^{3}\sin ^{4}\frac{\theta }{2}}\frac{1}{(1+\tau )}
\nonumber \\
&&\times \Biggl(G_{E}^{2}+\tau G_{M}^{2}+2\tau (1+\tau )G_{M}^{2}\tan ^{2}%
\frac{\theta }{2}  \nonumber \\
&&+h\left[ \frac{\epsilon _{1}+\epsilon _{3}}{M}\sqrt{\tau (1+\tau )}%
G_{M}^{2}\tan ^{2}\frac{\theta }{2}\,{\bbox{\zeta }}_{4}\cdot \widehat{{\bf z%
}}-2\sqrt{\tau (1+\tau )}G_{M}G_{E}\tan \frac{\theta }{2}\,{\bbox{\zeta }}%
_{4}\cdot \widehat{{\bf x}}\right] \Biggr)  \label{xsec}
\end{eqnarray}
where we take the unit vector $\widehat{{\bf z}}$ in the direction of ${\bf p%
}_{4}$, the unit vector $\widehat{{\bf y}}$ in the direction of ${\bf p}%
_{1}\times {\bf p}_{3}$ (i.e., perpendicular to the scattering plane) and
the unit vector $\widehat{{\bf x}}$ in the scattering plane and defined by $%
\widehat{{\bf x}}=\widehat{{\bf y}}\times \widehat{{\bf z}}$.

In (\ref{xsec}), the spin-independent terms give the well-known Rosenbluth
cross section. The remaining terms determine the longitudinal and
perpendicular polarization of the recoil proton \cite{akh}.

\section{Radiative Corrections to Elastic Electron-Proton Scattering}

In this section we consider each of the terms contributing to the radiative
correction to elastic electron-proton scattering as treated in the generally
used analysis given in \cite{tsai} and \cite{mo}. We show that if one makes
the approximations which are inherent to the derivation given in these
references, then the radiative correction to elastic electron-proton
scattering is the same for polarized and unpolarized electrons and protons.

The radiative correction is comprised of the purely elastic amplitudes
(electron and proton vertex corrections, electron and proton self energies,
box and crossed box diagrams, and vacuum polarization terms) and inelastic
amplitudes (emission of soft bremsstrahlung photons by any of the charged
particles). Let us consider each of these in turn. The cross section for
emission of soft photons, $d\sigma _{\text{brem}}$, is simply equal to a factor
which multiplies the one-photon exchange cross section, $d\sigma $, and that
factor is independent of the spins of the electrons and protons: 
\begin{equation}
d\sigma _{\text{brem}}=-\frac{\alpha }{4\pi ^{2}}d\sigma \int^{\prime }\frac{%
d^{3}k}{\omega }\left( \frac{p_{3}}{p_{3}\cdot k}-\frac{p_{1}}{p_{1}\cdot k}-%
\frac{p_{4}}{p_{4}\cdot k}+\frac{p_{2}}{p_{2}\cdot k}\right) ^{2}.
\label{brems}
\end{equation}

Consider next the radiative corrections to the purely elastic cross section.
To lowest order in $\alpha $ these are found from the cross product of the
matrix element for one-photon exchange, $M_{0}$, and the matrix elements for
each of the higher order processes: 
\begin{equation}
\left| {\cal M}\right| ^{2}=\left| M_{0}\right| ^{2}+2\text{Re}\left\{
M_{0}^{\dagger }\left( M_{1}+M_{2}+...\right) \right\} .  \label{rcmtrx}
\end{equation}
Thus, provided the matrix elements $M_{1},M_{2},$... can be expressed as $%
M_{0}$ times a factor which is independent of the spin of the particles, the
radiative correction for elastic scattering will factor as a
spin-independent term.

The matrix element for vacuum polarization, $M_{1}$, is, after charge
renormalization, related simply to the matrix element $M_{0}$ by 
\begin{equation}
M_{1}=M_{0}\sum_{i}\Pi (q^{2}/m_{i}^{2})  \label{vp}
\end{equation}
in which $\Pi (q^{2}/m_{i}^{2})$ is independent of the spins of the
particles \cite{tsai2}, \cite{bj} and the sum is carried over the electron
and higher mass particle-antiparticle loops.

The matrix element for the electron vertex correction, $M_{2},$ is given by 
\begin{equation}
M_{2}=e^{2}\overline{u}(p_{3})\Lambda ^{\mu }(p_{3},p_{1})u(p_{1})\frac{(-i)%
}{q^{2}+i\epsilon }\overline{u}(p_{4})\Gamma _{\mu }(q^{2})u(p_{2})
\label{M2}
\end{equation}
where 
\begin{equation}
\Lambda ^{\mu }(p_{3},p_{1})=-ie^{2}\int \frac{d^{4}k}{(2\pi )^{4}}\frac{1}{%
k^{2}-\lambda ^{2}+i\epsilon }\gamma ^{\nu }\frac{1}{(p\!\!\!/_{3}-k\!\!%
\!/-m+i\epsilon )}\gamma ^{\mu }\frac{1}{(p\!\!\!/_{1}-k\!\!\!/-m+i\epsilon )%
}\gamma _{\nu }.  \label{A5}
\end{equation}
Comparing (\ref{M2}) with (\ref{M0}) we see that if the spin-operator
dependence in $\Lambda ^{\mu }(p_{3},p_{1})$ reduces to $\gamma ^{\mu }$,
then $M_{2}$ will be a multiple of $M_{0}$, the factor being independent of
the spins of the particles. As it stands, the integral for $\Lambda ^{\mu
}(p_{3},p_{1})$ is divergent. However, if we introduce a convergence factor, 
$-\Lambda ^{2}/(k^{2}-\Lambda ^{2}+i\epsilon )$, in the integrand then the
integration can be carried out, and, taking the limit $\Lambda \rightarrow
\infty $, we find that $\Lambda ^{\mu }(p_{3},p_{1})$ has the form $%
G_{1}(q^{2})\gamma ^{\mu }+G_{2}(q^{2})\frac{i\sigma ^{\mu \nu }q_{\nu }}{2m}
$, where 
\begin{equation}
G_{1}^{(e)}(q^{2})=\frac{\alpha }{4\pi }\left\{ -2(2m^{2}-q^{2})\phi
_{1}(\lambda ^{2})+\left( \frac{3\rho ^{2}-4m^{2}}{\rho \rho _{1}}\right)
\ln x+\frac{1}{2}+\ln \left( \frac{\Lambda ^{2}}{m^{2}}\right) \right\}
\end{equation}
and 
\begin{equation}
G_{2}^{(e)}(q^{2})=\frac{\alpha }{4\pi }\left\{ \frac{4m^{2}}{\rho \rho _{1}}%
\ln x\right\}
\end{equation}
in which 
\begin{eqnarray}
\phi _{1}(\lambda ^{2}) &\,_{\overrightarrow{\lambda \rightarrow 0}}\,&\frac{%
1}{\rho \rho _{1}}\left\{ -2L\left( -\frac{1}{x}\right) -\frac{\pi ^{2}}{6}-%
\frac{1}{2}\ln ^{2}x+\ln x\ln \left( \frac{\rho ^{2}}{\lambda ^{2}}\right)
\right\} ,  \nonumber \\
&&L(z)=-\int_{0}^{z}\frac{\ln (1-t)}{t}dt,  \label{phi1lam}
\end{eqnarray}
with $\rho ^{2}=-q^{2}+4m^{2}$, $\rho _{1}^{2}=-q^{2}$, and $x=(\rho +\rho
_{1})/(\rho -\rho _{1})=(\rho +\rho _{1})^{2}/4m^{2}$. Thus for $%
-q^{2}>>m^{2}$ the term $G_{2}(q^{2})$ is of order $m^{2}/(-q^{2})$ relative
to $G_{1}(q^{2})$ and hence may be neglected, so that we have $M_{2}=$ $%
G_{1}(q^{2})M_{0}$. The inclusion of the self energy contribution for the
electron is obtained by subtracting $\Lambda ^{\mu }(p_{1},p_{1})$ from the
expression given in (\ref{A5}), giving 
\begin{equation}
\widetilde{M}_{2}=\left[ G_{1}(q^{2})-G_{1}(0)\right] M_{0}  \label{evtx}
\end{equation}
where, for $-q^{2}>>m^{2},$ 
\begin{eqnarray}
\quad G_{1}(q^{2})-G_{1}(0) & = & \frac{\alpha }{2\pi } \left\{ -\frac{1}{2}
\ln ^{2}\left( \frac{-q^{2}} {m^{2}}\right)\right. +\frac{\pi ^{2}}{6} 
\nonumber \\
& &-\left[ \ln \left( \frac{-q^{2}}{m^{2}}\right) -1\right] \ln \left( \frac{%
m^{2}}{\lambda ^{2}}\right)  \nonumber \\
& & +\frac{3}{2}\left. \ln \left( \frac{-q^{2}}{m^{2}}\right) -2 \right\}.
\end{eqnarray}

Finally, we consider the proton vertex correction and the box and crossed
box contributions, $M_{3}$, $M_{4},$ and $M_{5}$, respectively. The matrix
elements for these corrections are given by 
\begin{equation}
M_{3}=e^{2}\overline{u}(p_{3})\gamma ^{\mu }u(p_{1})\frac{(-i)}{%
q^{2}+i\epsilon }\overline{u}(p_{4})\Lambda _{\mu }(p_{4},p_{2})u(p_{2})
\label{A4}
\end{equation}
with 
\begin{eqnarray}
\Lambda _{\mu }(p_{4},p_{2}) &=&-ie^{2}\int \frac{d^{4}k}{(2\pi )^{4}}\frac{1%
}{k^{2}-\lambda ^{2}+i\epsilon }\Gamma ^{\nu }(k^{2})\frac{1}{%
(p\!\!\!/_{4}-k\!\!\!/-M+i\epsilon )}\Gamma _{\mu }(q^{2})  \nonumber \\
&&\times \frac{1}{(p\!\!\!/_{2}-k\!\!\!/-M+i\epsilon )}\Gamma _{\nu }(k^{2}),
\label{A6}
\end{eqnarray}
\begin{eqnarray}
M_{4} &=&(e^{2})^{2}\int \frac{d^{4}k}{(2\pi )^{4}}\frac{1}{k^{2}-\lambda
^{2}+i\epsilon }\frac{1}{(k-q)^{2}-\lambda ^{2}+i\epsilon }  \nonumber \\
&&\times \left[ \overline{u}(p_{3})\gamma ^{\nu }\frac{1}{%
p\!\!\!/_{1}-k\!\!\!/-m+i\epsilon }\gamma ^{\mu }u(p_{1})\right]  \label{box}
\\
&&\times \left[ \overline{u}(p_{4})\Gamma _{\nu }((k-q)^{2})\frac{1}{%
p\!\!\!/_{2}+k\!\!\!/-M+i\epsilon }\Gamma _{\mu }(k^{2})u(p_{2})\right] , 
\nonumber
\end{eqnarray}
and 
\begin{eqnarray}
M_{5} &=&(e^{2})^{2}\int \frac{d^{4}k}{(2\pi )^{4}}\frac{1}{k^{2}-\lambda
^{2}+i\epsilon }\frac{1}{(k-q)^{2}-\lambda ^{2}+i\epsilon }  \nonumber \\
&&\times \left[ \overline{u}(p_{3})\gamma ^{\nu }\frac{1}{%
p\!\!\!/_{1}-k\!\!\!/-m+i\epsilon }\gamma ^{\mu }u(p_{1})\right]  \label{cbx}
\\
&&\times \left[ \overline{u}(p_{4})\Gamma _{\mu }(k^{2})\frac{1}{%
p\!\!\!/_{4}-k\!\!\!/-M+i\epsilon }\Gamma _{\nu }((k-q)^{2})u(p_{2})\right] .
\nonumber
\end{eqnarray}
In general, these matrix elements depend on the initial and final spin
states, and are not proportional to $M_{0}$ times a spin independent factor.

Now consider the approximation used in \cite{tsai} to evaluate these matrix
elements, which we call here the soft-photon approximation. The integrands
in $M_{4}$ and $M_{5}$ have two infrared divergent factors, $[(k^{2}-\lambda
^{2}+i\epsilon )((k-q)^{2}-\lambda ^{2}+i\epsilon )]^{-1}$, and are thus
peaked when either of the two exchanged photons is soft, becoming divergent
when $k\rightarrow 0$ or when $k\rightarrow q$. We therefore first
rationalize the propagators so that all spin matrices are in the numerator
and then evaluate the {\it numerators} in $M_{4}$ and $M_{5}$ at these two
points (first setting $k=0$ and then setting $k=q$; note that $\Gamma _{\mu
}(0)=\gamma _{\mu }$) but make no changes to the denominators. A simple
calculation using the fact that we have on-shell particles shows that in
fact each of the numerators has the same value for $k=0$ as for $k=q$, {\it %
viz}., $4ip_{1}\cdot p_{2}q^{2}M_{0}$ in the case of $M_{4}$ and $%
4ip_{3}\cdot p_{2}q^{2}M_{0}$ in the case of $M_{5}$. Taking this factor
outside of the integral, we are left with a scalar four-point function,
independent of the particle spins: 
\begin{eqnarray}
M_{4} &=&8ie^{2}q^{2}M_{0}p_{1}\cdot p_{2}\int \frac{d^{4}k}{(2\pi )^{4}}%
\frac{1}{k^{2}-\lambda ^{2}+i\epsilon }\frac{1}{(k-q)^{2}-\lambda
^{2}+i\epsilon }  \label{sphbox} \\
&&\times \frac{1}{(k^{2}-2k\cdot p_{1}+i\epsilon )}\frac{1}{(k^{2}+2k\cdot
p_{2}+i\epsilon )}  \nonumber
\end{eqnarray}
and 
\begin{eqnarray}
M_{5} &=&8ie^{2}q^{2}M_{0}p_{3}\cdot p_{2}\int \frac{d^{4}k}{(2\pi )^{4}}%
\frac{1}{k^{2}-\lambda ^{2}+i\epsilon }\frac{1}{(k-q)^{2}-\lambda
^{2}+i\epsilon }  \label{sphcbx} \\
&&\times \frac{1}{(k^{2}-2k\cdot p_{1}+i\epsilon )}\frac{1}{(k^{2}-2k\cdot
p_{4}+i\epsilon )}.  \nonumber
\end{eqnarray}
We note that in \cite{tsai} an approximation is also made in the
denominators of these integrals, reducing these four-point functions to
three-point functions, but this is not needed for the conclusions of the
present paper.

In the case of $M_{3}$, the integrand is peaked when $k=0$; we therefore set 
$k=0$ in all terms of the {\it numerator }of $M_{3}$, again using the fact
that we have on-shell particles, and find

\begin{equation}
M_{3}=-4ie^{2}p_{4}\cdot p_{2}M_{0}\int \frac{d^{4}k}{(2\pi )^{4}}\frac{1}{%
(k^{2}-\lambda ^{2}+i\epsilon )}\frac{1}{(k^{2}-2k\cdot p_{4}+i\epsilon )}%
\frac{1}{(k^{2}-2k\cdot p_{2}+i\epsilon )}.  \label{pvtx}
\end{equation}
With the soft-photon approximation, the proton vertex correction is a
multiple of $M_{0}$, and, as with $M_{2}$, the factor is independent of the
spins of the particles. Again because of the soft-photon approximation, the
self-energy contribution is essentially the same as that obtained for the
electron: since the virtual photon in the self-energy diagrams is assumed to
be soft, its interaction with the proton is given by $\gamma _{\mu },$ as in
the case of the electron, so that the self-energy contribution is obtained
by subtracting $\Lambda _{\mu }(p_{2},p_{2})$ from the expression given in (%
\ref{A6}).

Thus, substituting the expressions for $M_{1},$ $M_{2},$ $M_{3},$ $M_{4}$
and $M_{5}$ given in (\ref{vp}), (\ref{evtx}), (\ref{pvtx}), (\ref{sphbox}),
(\ref{sphcbx}) in (\ref{rcmtrx}) and adding the contribution from real soft
photons (\ref{brems}), we see that the cross section can be written in the
form:
\begin{equation}
d\sigma _{\text{corr}}=d\sigma \left( 1+\delta \right) .
\end{equation}
in which the radiative correction term $\delta $ is independent of the spins
of the particles.

\begin {acknowledgments}
It is a pleasure to acknowledge stimulating conversations between M. Gar\c con
and one of the authors (L.C.M.), which provided the impetus for this work.
\end {acknowledgments}

\end{document}